# Thermal treatment effects on PMN-$_{0.4}$PT/Fe multiferroic heterostructures

*Deepak Dagur, Alice Margherita Finardi, Vincent Polewczyk, Aleksandr Yu. Petrov, Simone Dolabella, Federico Motti, Hemanita Sharma, Edvard Dobovicnik, Andrea Giugni, Giorgio Rossi, Claudia Fasolato, Piero Torelli, and Giovanni Vinai\**

D. Dagur, A. M. Finardi, V. Polewczyk, A. Yu. Petrov, S. Dolabella, F. Motti, H. Sharma, A. Giugni, G. Rossi, P. Torelli, G. Vinai

CNR-IOM - Istituto Officina dei Materiali, S.S. 14 km 163.5, I-34149, Trieste, Italy

E-mail: vinai@iom.cnr.it

D. Dagur, H. Sharma

Department of Physics, University of Trieste, Trieste, Via Alfonso Valerio 2, 34127, Italy

E. Dobovicnik

Elettra - Sincrotrone Trieste S.C.p.A., Strada Statale 14 km 163.5, 34149 Trieste, Italy

A. M. Finardi, A. Giugni, G. Rossi

Department of Physics, University of Milan, Milan, Via Festa del Perdono 7, 20122, Italy

C. Fasolato

Istituto dei Sistemi Complessi (ISC)-CNR, Rome, Piazzale Aldo Moro 5, 00185, Italy

**ABSTRACT**

Multiferroic heterostructures have gained in recent years a renewed role in spintronic applications due to their possibility in controlling the magnetic properties via interfacial coupling by exploiting the ferroelectric response to various external stimuli. Whereas the main mechanisms ruling the converse magnetoelectric coupling are considered as established, the research on how to optimize the ferroelectric properties is an active field. In particular the complex phase diagram of [Pb(Mg$_{1/3}$Nb$_{2/3}$)O$_3$]-$_x$[PbTiO$_3$] (PMN-$_x$PT) single crystals, that present relaxor ferroelectric and photovoltaic properties, deserves further investigation. For instance, crystalline quality and thermal stability of the ferroelectric domains in heterostructures need assessment. Here we show how, by thermal annealing over the ferroelectric Curie temperature and then cooling PMN-$_{0.4}$PT/Fe heterostructures in inert atmosphere the domain population is significantly modified, evolving from a highly disordered, mostly out-of-plane domain population, to improved crystallinity and prevalent in-plane oriented domains. Upon further annealing, the domain population switches back to prevalent out-of-plane orientation suggesting that intermediate annealing steps can freeze PMN-$_{0.4}$PT domain population in a metastable configuration. The structural analysis was carried out by combining micro-Raman and X-ray diffraction (XRD) measurements. In the three states the magnetic properties of interfacial Fe thin film are affected by the ferroelectric configurations as a consequence of changing interfacial strain, evolving from an isotropic behavior to an anisotropic one and back. These results are addressed further investigations, on both micro and macroscopic scales, on the domain population and thermal stability in ferroelectric crystals, on how structural optimization affects the global and local ferroelectric polarization, and finally on their interfacial coupling with magnetic layers.

# 1. INTRODUCTION

In multiferroic (MF) heterostructures may coexist two or more ferroic order parameters coupled at the interface, such as ferromagnetic (FM), ferroelectric (FE), ferroelastic or ferrotoroidic (FTO).[1,2] Among these four orders, the most investigated ones are the first two, with a particular focus on the possibility of modifying the FM (break in time-reversal symmetry) and FE (break in space-inversion symmetry) orders via direct or converse magnetoelectric (ME) coupling at their interface. This offers a wealth of opportunity for technological implementations in case of converse ME coupling, i.e., modifying the FM response by inducing changes on the FE layer. The ME coupling is mediated by five main mechanisms: charge coupling, ion migration, exchange bias, orbital, and strain-transfer effects across the interface.[3,4] More recently, additional parameters such as fully optical processes,[5–7] and reversible morphological modifications upon electric polarization[8,9] proved to be valuable additional levers for tuning the interfacial properties in MF heterostructures.

In this context, the optimization of the FE properties, with the close link between structural and ferroelectric ones, is a key aspect to maximize the modifications of the interfacial FM layer in MF heterostructure. An effective manipulation of FE domain size and polarization direction can substantially modify the piezoelectric, dielectric, electromechanical and optical properties of the FE layer, and therefore the FM one. Many methods have been reported in last few years to modify

FE domain structures and to play with domain wall engineering, such as electric polarization[10–14] or visible light induced photostriction[5,15–19]. While these methods act on pristine substrates, another possible approach is to modify the structural properties via thermal treatments,[20–25] specifically by crossing the FE Curie temperature $T_C$. The transition from FE domains to a cubic structure, i.e. in the centrosymmetric paraelectric phase (PE) and backward upon cooling determines a reorganization of the FE domains, and can lead to modifications on the concentration of local defects and to FE domain reorientation.[32–35]

Among FE single crystals, $Pb(Mg_{1/3}Nb_{2/3})O_3$-$_x PbTiO_3$ (PMN-$_x$PT) gained a privileged position due to its large piezoelectric coefficient and relaxor properties,[29–32] whereas degradation of ferroic properties and aging upon several polarization switching[33,34] and its disordered relaxor nature[35] leave room for further optimizations.

Here we report the FE structural modifications of PMN-$_{0.4}$PT (001) substrates upon annealing above $T_C$ and how they induce modifications on the magnetic properties of an interfacial Fe thin film. We show via XRD and micro-Raman characterizations how the FE domain population is modified upon subsequent thermal annealing cycles, shifting from a prevalent out-of-plane polarized state to a prevalent in-plane one, along with an improvement of the crystalline quality. These changes trigger the modification of the FM properties of the interfacially coupled Fe thin film. By means of magneto-optic Kerr effect (MOKE) characterization, we show how the thermal treatments affect both the anisotropy and the magnetic coercive field ($H_C$) of the Fe film. These experimental results demonstrate how thermal treatments are an additional lever for modifying the structural and interface properties of multiferroic heterostructures at both the macro and microscale with direct implications on the correlation between FE polar properties and domain orientations and FM properties.

## 2. EXPERIMENTAL SECTION

**2.1. Sample preparation.** Pristine one-sided polished (001) PMN-$_{0.4}$PT 2.5 × 2.5 mm$^2$ substrates with thickness of 0.5 mm from SurfaceNet GmbH supplier have been used as FE substrates. Substrates were cleaned with standard acetone and ethanol procedures in ultrasonic bath and rinsed with N$_2$ flow. The substrates were then introduced in the growth chamber at the MBE Cluster of NFFA-Trieste facility,[36] where 4 nm thick films of Fe were deposited at room temperature, base pressure 3 × 10$^{-10}$ mbar, with Fe deposition rate calibrated using quartz microbalance and fixed at 1.5 Å min$^{-1}$. Samples were then capped by a 5 nm thick MgO layer with 0.5 Å min$^{-1}$ deposition rate. After deposition, the thermal treatments on the heterostructures were done inside a N$_2$ filled glovebox (MBraun, O$_2$, H$_2$O < 0.5 ppm).

**2.2. XAS characterizations.** X-ray absorption spectra were carried out at APE-HE beamline of NFFA at the Elettra synchrotron radiation facility in Trieste.[37] The spectra were measured at room temperature in total electron yield (TEY) mode, normalizing the intensity of sample current to the incident photon flux current at each energy value. The sample surface was kept at 45° with respect to the incident x-rays, for a beam footprint on the sample surface of around 200 µm large.

**2.3. Structural XRD Characterizations.** XRD measurements were performed with a four-circle X'Pert PANalytical diffractometer available at the NFFA-Trieste facility, using a monochromatic Cu-K$\alpha_1$ radiation ($\lambda$ = 1.54056 Å) in the Bragg-Brentano configuration at room temperature. The scans were taken in high-resolution mode with incident optics composed of a 4-bounce Ge(220) monochromator to select only Cu-K $\alpha_1$ line.

**2.4. Micro-Raman characterizations.** Raman spectra were collected in the backscattering geometry using a Horiba (LabRAM HR-Evolution) Raman microscope equipped with a He-Ne laser of 632.8 nm wavelength yielding the output power of 20 mW (see Figure 4a). The beam passes through a polarizer ($P_1$), to eliminate spurious depolarizing effects, and a lambda halfwave plate ($R_1$). An additional polarization rotator ($R_s$) is placed on the microscope head to rotate the incoming polarization of a given angle φ, and to rotate the scattered light backwards of the same angle. From a symmetry point of view, this element is equivalent to a rotation of the sample about the microscope axis. The beam passes through a 20x microscope lens with 3 mm focal length, for beam spot size of approximately 3 μm. In the backscattering geometry detection mode, the inelastically scattered signal is collected by the same objective and passes through the notch filter. The signal is then guided by a set of mirrors to a dispersive spectrometer.

Before entering into the spectrometer, a second polarizer ($P_2$) was used to select the scattered radiation with components of the polarization either parallel or perpendicular to the polarization of the excitation beam, followed by a second lambda halfwave plate ($R_2$) to rotate the light polarization backwards, as so to enter the spectrometer with a fixed polarization configuration. The spectrometer (800 mm focal length) was equipped with a 600 grooves/mm diffraction grating, providing a high spectral resolution almost equal to $3\pm1$ cm$^{-1}$. Rayleigh scattered light was removed by three BragGrate Notch filters, allowing to collect the Raman signal down to very low wavenumbers (about 10 cm$^{-1}$). Finally, a Peltier cooled CCD was used to collect the Raman scattered light signals.

**2.5. MOKE measurements.** Magnetic hysteresis curves have been taken with longitudinal MOKE measurements at the MBE Cluster of NFFA-Trieste facility.[36], using an *s*-polarized red laser (658 nm wavelength). The laser spot size on the sample was estimated around 500 μm. The coercive field was evaluated after averaging the half width hysteresis loops at zero magnetization for both negative and positive sides of the loop.

## 3. RESULTS AND DISCUSSION

### 3.1. PMN-$_{0.4}$PT/Fe thermal treatments

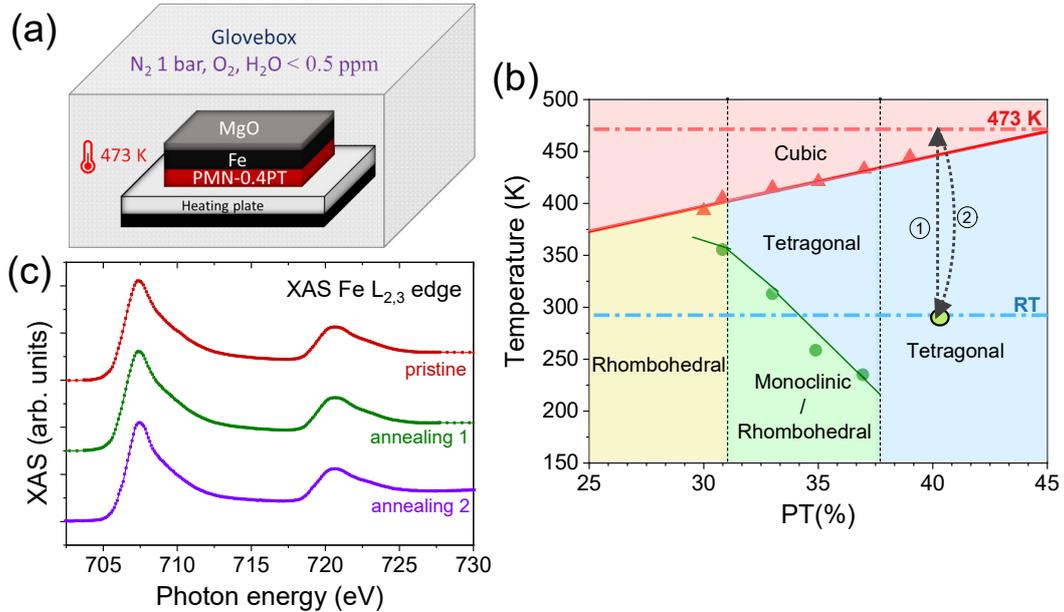

**Figure 1. (a)** Schematics of thermal treatment setup of PMN-$_{0.4}$PT/Fe heterostructure with heating in glovebox; **(b)** PMN-$_x$PT phase diagram[38]: arrows indicate the thermal treatment process, going from room temperature (RT) up to 473 K, crossing the structural transition (1), then back to RT (2). Horizontal dashed lines indicate the two critical temperatures of the thermal treatment; **(c)** X-ray absorption spectra at Fe $L_{2,3}$ edges for the pristine unannealed and after the two thermal treatments for PMN-$_{0.4}$PT/Fe heterostructure.

Pristine PMN-$_{0.4}$PT/Fe MF heterostructures have been grown at the MBE Cluster of NFFA-Trieste facility.[36] After deposition and characterizations of the pristine heterostructure, the samples and pristine PMN-$_{0.4}$PT substrates underwent thermal treatments inside a N$_2$ filled glovebox to minimize interactions with oxygen during the process. The heterostructures were heated on a heating plate (heating ramp 6 K/min, cooling ramp ~3 K/min) up to 473 K, *i.e.* over T$_C$ for PMN-$_{0.4}$PT.[38,39] **(Figure 1)**. The same process was repeated twice on the same heterostructure, with different heating time span, 15 minutes during the first one (from now on, annealing 1) and 180 minutes in the second one (annealing 2). X-ray absorption spectroscopy measurements at Fe $L_{2,3}$ edges, taken at APE-HE beamline of NFFA at the Elettra synchrotron radiation facility in Trieste[36], showed no chemical modifications of the Fe thin film, in particular no sign of oxidation coming either from the PMN-$_{0.4}$PT substrate or from the atmosphere, even after air exposure, excluding any thermally induced interfacial intermixing or oxygen migration between Fe and PMN-$_{0.4}$PT during the annealing processes (Figure 1c).

**3.2. PMN-$_{0.4}$PT thermally driven structural modifications: X-ray diffraction**

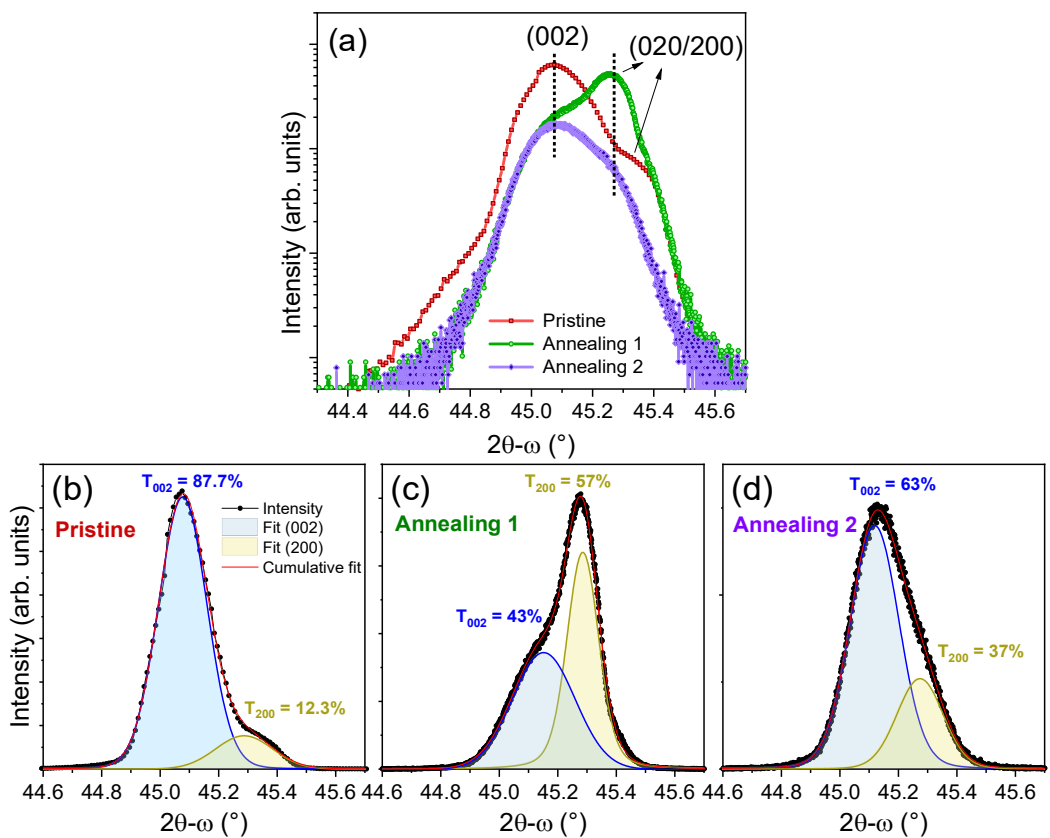

**Figure 2. (a)** XRD 2θ-ω scans for pristine unannealed and after the two thermal treatments for bare PMN-$_{0.4}$PT (001) substrate; **(b-d)** fitting of the (002) and (200)/(020) contributions for the three cases.

**Figure 2** shows the symmetric XRD 2θ-ω scans of (002) peak of the PMN-$_{0.4}$PT substrate for pristine and thermally treated cases. As shown in Figure 1b, the phase diagram of PMN-$_{0.4}$PT presents at room temperature three structural regimes according to PT concentration (dashed light blue line), going from rhombohedral to tetragonal, with an intermediate range in which monoclinic and rhombohedral structures coexist. In case of PMN-$_{0.4}$PT, this PT concentration leads to a purely tetragonal structure. In the pristine case (red curve), *i.e.* before any thermal treatment, the profile shows two main peaks of different intensities, one corresponding to out-of-plane (002) FE domains and the second one to in-plane (200/020) domains, with the former being more intense than the latter. From these peaks, it is possible to evaluate the lattice parameters of the tetragonal structure, being ($c$)$_{tet}$ for (002) at 45.07° equal to (4.022 ± 0.001) Å and ($a=b$)$_{tet}$ for (002/020) at 45.25° equal to (3.9936 ± 0.001) Å, in good agreement with what reported in literature.[40,41] After the first thermal treatment, the intensity of the two peaks is reversed, with the in-plane component (200/020) becoming predominant as compared to the out-of-plane one (002). In this case, the lattice parameters were found to be 4.0201 and 4.0025 Å, respectively, *i.e.* a reduction of $c$ and increase of $a$ lattice parameter. After the additional 180' annealing, the in-plane component becomes negligible, with the presence of a single dominant peak at 45.14° (*i.e.* $c$ = 4.0146 Å), which we attribute to an out-of-plane (002) orientation of the FE domains.

By fitting the relative intensities of the two peaks for the three cases (Figure 2b-d), we obtained the 87.7% of out-of-plane domains and 12.3% of in-plane one for the pristine unannealed case, while the percentages move to 43% and 57% after annealing 1 treatment, indicating a substantial increase in the population of in-plane domains due to quick thermal treatment, whose variability appears to be sensitive to the annealing time. After an additional 180' of annealing, we observed the return to a majority of domains to out-of-plane direction (63%). A similar evolution of the ratio between in-plane and out-of-plane domains was confirmed for PMN-$_{0.4}$PT/Fe substrate (Figure S1, supporting information), going from 88% and 12% for out-of-plane and in-plane domains in case of pristine unannealed case to 40% and 60% for annealing 1. This proves a good reproducibility and a negligible influence of the interfacial layer on their thermal evolution.

After the symmetric scans, we carried out the 2-dimensional XRD reciprocal space maps (2D-RSMs) recorded asymmetrically around the (103) reflection. **Figure 3** shows the RSMs for the bare PMN-$_{0.4}$PT heterostructure in the pristine unannealed and two different time spans of annealing cases.

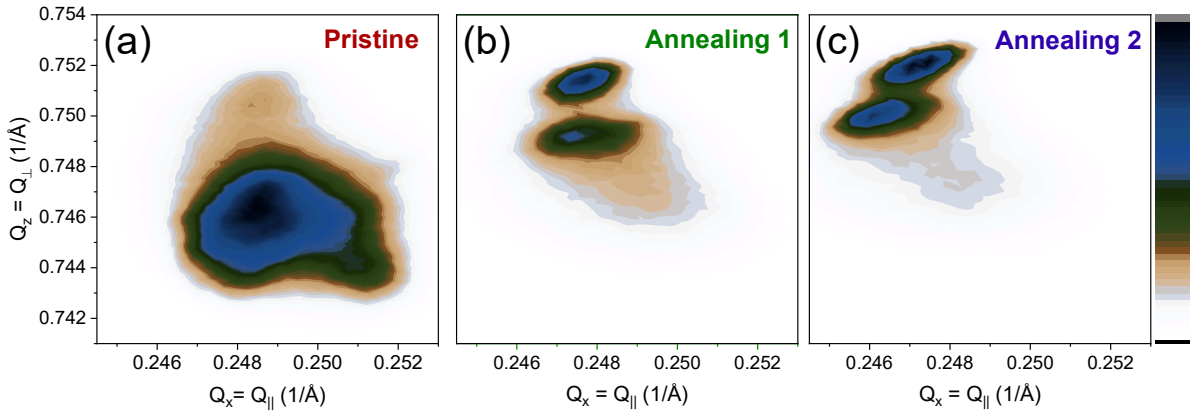

**Figure 3.** 2D-RSMs of the (103) Bragg peak of PMN-$_{0.4}$PT (001) substrate for **(a)** pristine unannealed and after thermal treatments, *i.e.* **(b)** annealing 1 and **(c)** annealing 2.

The reciprocal lattice point (103) of the pristine unannealed sample shows an inhomogeneous distribution of the diffracted intensities in both $Q_x$ and $Q_z$ directions of the reciprocal space. Furthermore, the diffuse x-ray scattering around the maxima reveals a high level of structural disorder, which can be attributed to the presence of lattice defects such as dislocations and clusters of point defects. The latter are often associated with the coexistence of polar nanoregions and chemically ordered clusters, leading to intrinsic short- and long-range disorder.[42–44]

The two processes of thermal annealing of PMN-$_{0.4}$PT produce different patterns of the diffracted intensity distribution. Firstly, we can notice sharper and more defined maxima in correspondence to the two domain distributions, indicating the formation of two well separated crystalline domains.

### 3.3. micro-Raman PMN-$_{0.4}$PT characterization

Micro-Raman spectra were collected on pristine and thermally treated PMN-$_{0.4}$PT, in the spectral range from 10 to 410 cm$^{-1}$ for a probed area of around 3 μm$^2$. XRD scans of the samples used for micro-Raman characterizations are shown in Figure S2, supporting information. The schematics of the setup and three different optical configurations for measuring the spectra is shown in Figure 4a and 4b, respectively, whereas details can be found in the experimental section. Figure 4c shows an example of a spectrum acquired on the pristine unannealed sample, along with the fitting curve and deconvoluted Raman modes, obtained by combining eight Lorentzian peaks in correspondence to the main features of the spectrum. An index 'Ai' is attributed to each peak, as shown in the figure. Peaks labelled from A2-A8 are associated with the vibrational modes of the system, while A1 is attributed to a residual elastic scattering component.

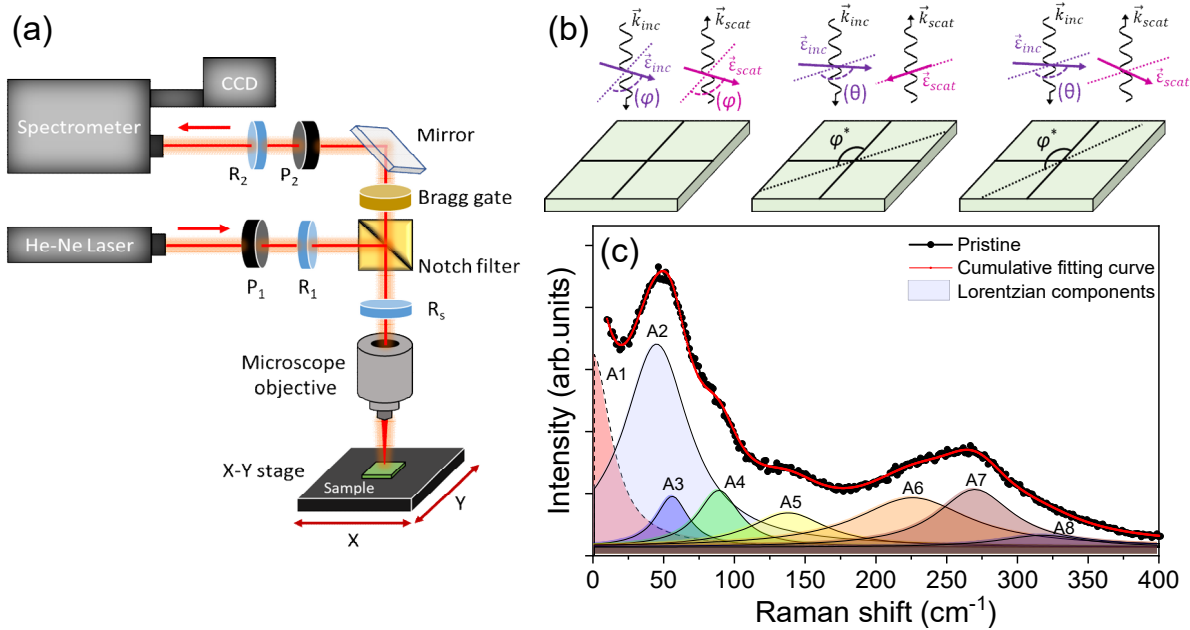

**Figure 4: (a)** Schematics of the micro-Raman setup used for polarization-resolved measurements in backscattering mode. The polarizers and halfwave plate retarders are indicated by $P_i$ and $R_i$, respectively. **(b)** Schematic of the three used polarization configurations. **(c)** example of a Raman spectrum with the fitted Raman modes A2-A8 and fitting curve (red).

The Raman shifts of the fitted peaks are consistent with what reported in literature.[45] The modes at lower frequencies (A2 at 43 cm$^{-1}$ and A3 at 50 cm$^{-1}$) are attributed to the lifting of degeneracy of the low wavenumber modes of PMN, associated to the variable elemental composition of the B site of PMN-$_{0.4}$PT. The other peaks are attributed to lattice modes (A4 at 88 cm$^{-1}$ and A5 at 137 cm$^{-1}$), and oxygen vibrations (A6 at 220 cm$^{-1}$) and A7 at 266 cm$^{-1}$).

Micro-Raman spectra were collected by combining the incident and scattered light polarizations as a function of the sample crystallographic orientation and of the relative orientation between polarization and sample axes.

The spectra were observed for three different configurations (as already shown in Figure 4b):

a) the polarization of both incident and scattered beams was set to vertical; the spectra collected for the polar plot were taken while rotating the crystallographic sample direction φ ($\varepsilon_{inc} = (\varphi)$, $\varepsilon_{scat} = (\varphi)$);

b) the rotation of the sample axis was set at a selected **φ\*** value, corresponding to a high symmetry direction defined from the trend measured with a) configuration; the incident beam polarization was rotated by a variable angle θ, while the scattered one was set to vertical, *i.e.* parallel to **φ\*** direction ($\varepsilon_{inc} = (\varphi) + \boldsymbol{\varphi}^*$, $\varepsilon_{scat} = \boldsymbol{\varphi}^*$);

c) the rotation of the sample axis was set at a selected **φ\*** value; the incident beam polarization was rotated by a variable angle θ, while the scattered one was set to horizontal, *i.e.* orthogonal to *φ\** direction ($\varepsilon_{inc} = (\varphi) + \boldsymbol{\varphi}^*$, $\varepsilon_{scat} = 90° + \boldsymbol{\varphi}^*$).

The combination of these three sets of angular dependent characterizations allows having a rather complete picture of the phononic modes present in the heterostructure, exploring all the elements of the polarizability tensor accessible in the adopted backscattering configuration.[46]

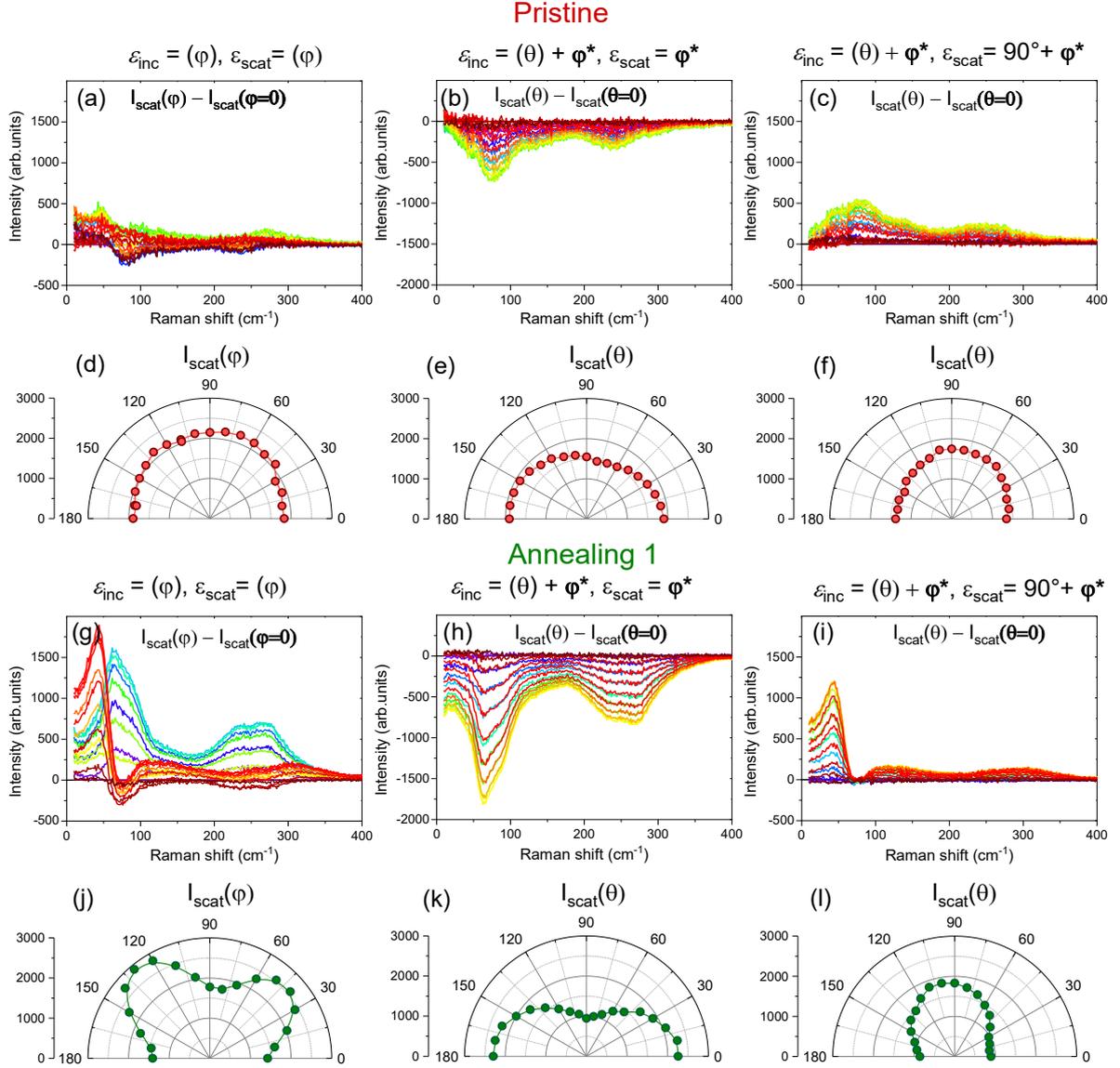

**Figure 5:** Polarization dependent Raman results obtained from PMN-$_{0.4}$PT/Fe heterostructure for **(top, a-f)** pristine and **(bottom, g-l)** after annealing 1 case. Figures **(a-c, g-i)** show the spectral variations in the three configurations, indicated on top of each panel by the direction of the incident and scattered polarizations (orientations that are varied, sweeping over 180° of rotation, are indicated in brackets). Polar plots **(d-f, j-l)** present the A3 mode intensity as a function of the rotating angle.

The full set of polarization dependent Raman spectra taken on PMN-$_{0.4}$PT/Fe (001) heterostructures in the three described configurations are shown in Figure S3, supporting information. To better underline the relative modifications of the Raman spectra as a function of the probed angles, we show in **Figure 5** the variations of the Raman spectra compared to a chosen angle (Figures 5a-c for pristine and 5g-i for annealed 1 cases), together with the polar plots as a function of the angle for the A3 lattice mode (see Figure 5d-f). In the case of pristine PMN-$_{0.4}$PT/Fe heterostructure, the polar plots show an almost perfectly isotropic response, signature of a lack of polarization dependence in the response of the phononic modes. This is an indication of a poorly defined crystalline order in the probed area, consistently with a large presence of local defects and an average randomly distribute lattice orientation in the three directions. This lack of feature variations is reported through the whole Raman spectrum, with differential curves which present faint features for the three optical configurations. On the other hand, once thermally treated, the increased crystalline quality is reflected on the local phononic response of the heterostructure. Differential curves present important variations, especially pronounced on the low Raman shift modes, as it can be seen in the differential curves of Figure 5g-i. Once plotted as a function of the intensity of specific lattice modes, the anisotropic behavior of the annealed sample becomes evident. Figure 5j-l show the polar plots for the A3 lattice mode. A well-defined 2-fold and 4-fold symmetries can be seen in the three configurations, indicating a well-defined symmetry of the mode, clearly responding to the polarization direction. Similar anisotropic angular dependences were observed also for lattice modes A2 and A5 and oxygen vibrational A7 mode (see Figure S4-S6, supporting information).

## 3.4. Thermal treatment effects on the magnetic properties of interfacial Fe thin film

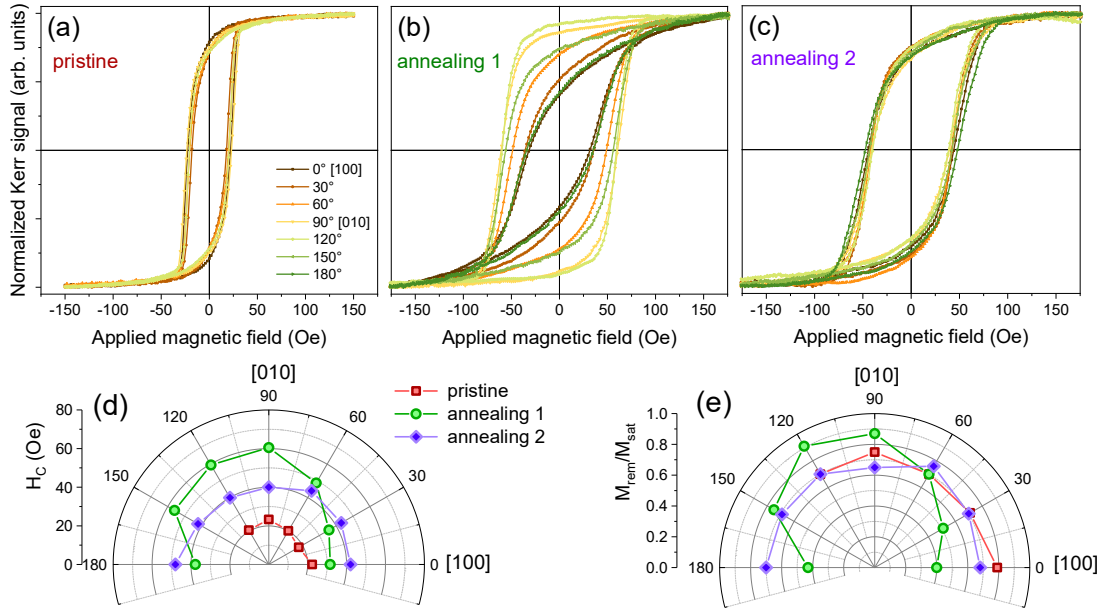

**Figure 6.** Angular dependent MOKE hysteresis loops for **(a)** pristine and **(b, c)** after thermal treatments, with corresponding polar plots of **(d)** coercive field and **(e)** magnetic remanence. All hysteresis loops are normalized to the magnetic saturation signal.

The magnetic properties of as grown PMN-$_{0.4}$PT/Fe heterostructure at room temperature were analyzed by MOKE measurements (**Figure 6**), see Figure S7, supporting information for the schematic representation of the setup.

In plane angular dependent hysteresis loops showed an almost isotropic behavior for the as-grown ''pristine'' sample (Figure 6a), with a coercive field of around 22 Oe along the [100] and 20.5 Oe along the [010] one. The measurements were then repeated after the first thermal treatment (Figure 6b). A drastic modification in the angular dependence is observed, passing from an isotropic behavior to an anisotropic one, with the easy axis laying along [010] axis and the hard one along [100]. This is consistent with the modification of the interfacial strain driven by the shift from a mostly out-of-plane to dominant in-plane PMN-PT domain population (see Figure S1, supporting information). Being PMN-$_{0.4}$PT tetragonal, in-plane crystallographic anisotropy becomes uniaxial in presence of ordered in-plane domains, while no preferential directions are present when most of the domains are out-of-plane, i.e. with $a = b$ lattice parameters at the interface with Fe. After the second thermal treatment, the isotropic magnetic response of the heterostructure confirms the strain-driven magnetic control induced by thermal treatments, since out-of-plane domains are again majoritarian (Figure 6c). The variation of coercive field compared to the pristine case (see polar plots, Figure 6d-e) can be attributed to the difference in crystallographic order between the two cases, while thermal-induced partial degradations or oxidations have been excluded via XAS measurements (Figure 1c).

## 4. CONCLUSIONS

The thermal evolution of ferroelectric and structural properties of PMN-PT single crystals has been reported in literature mostly in the morphotropic region. For instance, Katzke *et al.*[47] showed an hysteretic thermal dependent optical transmittance of PMN-$_{0.29}$PT, and attributed it to the different phase paths along the thermal cycles, with differences between unpoled and poled cases. Li *et al.*[48] showed different thermal expansions coefficients in thermal cycles crossing PMN-$_{0.31}$PT Curie temperature. In the PMN-$_{0.4}$PT case, Chien *et al.*[49] showed a tetragonal-to-cubic transition taking place starting from 450 K, with cubic regions expanding up to 465 K. Our results present the effects at room temperature after thermal cycles, showing a modification of both domain composition and defects concentration. Regarding this second aspect, their presence is linked to randomly polarized nanoregions (PNRs), which are indicated as the key features responsible of the piezoelectric and dielectric response of the monoclinic PMN-$_x$PT ferroelectric crystals.[35] The confirmation of PMN-$_{0.4}$PT to be a ferroelectric is shown in Figure S9, supporting information by means of characteristic I(E) curves.

Higher crystalline quality and reduced local defects become crucial when interfacially coupled with a FM layer, an aspect rarely investigated in literature. The possibility of modifying the FE domain homogeneity via thermal treatments has been up to now only reported by Y. Zhao *et al.*[50] on tetragonal PMN-PT machined tetragonal crystals, showing not only an elimination of disordered domains upon thermal cycles, but also crack-free crystals upon electric poling, a second aspect crucial and often neglected when coupling with FM thin films. Higher crystalline quality and reduced local defects become crucial when interfacially coupled with a FM layer, an aspect rarely investigated in literature. Better-defined FE domain dimensions and orientations are desirable for engineering interfacial coupling on domains with uniform properties and known orientation, as very recently shown by Ochoa *et al.* on $Fe_{75}Al_{25}$/$BaTiO_3$ heterostructures.[51] These aspects, combining morphological properties, ferroelectric domain composition and dimensions, and polarization properties upon electrical poling, stimulate for further investigations on interfacially coupled thermally treated heterostructures.

To conclude, in this work we have shown the effects of thermal treatment on PMN-$_{0.4}$PT/Fe multiferroic heterostructure by combining structural, spectroscopic and magnetic characterizations. Pristine unannealed PMN-$_{0.4}$PT substrate showed a combination of majoritarian out-of-plane and minoritarian in-plane domains, with a low crystallographic quality and isotropic lattice response, sign of a high local disorder and hinting at large presence of local defects. Once thermally treated by annealing over PMN-$_{0.4}$PT Curie temperature, the FE substrate showed a substantial increase of crystallographic quality and local order. Additionally, a switch of the domain population towards mostly in-plane direction was observed during the first annealing round, while out-of-plane domains returned to be the majority after subsequent annealing. These results present an additional electric field free possibility in switching the FE domain population and meanwhile modifying the magnetic anisotropy of the interfacial FM layer, whose potential engineering could be of high benefit once implemented in multiferroic-based devices.

ASSOCIATED CONTENT

**Supporting Information**.

The Supporting information is available free of charge.

Additional details about XRD characterizations of pristine and annealed PMN-$_{0.4}$PT/Fe heterostructures; polarization angular dependent micro-Raman measurements at A2, A5, and A7 vibrational modes; MOKE setup with angular dependent Ni hysteresis loops for short annealed sample; ferroelectric switching curve of PMN-$_{0.4}$PT/Fe.

AUTHOR INFORMATION

**Corresponding Author**
**Giovanni Vinai -** CNR-IOM - Istituto Officina dei Materiali, S.S. 14 km 163.5, I-34149, Trieste, Italy

E-mail: vinai@iom.cnr.it

**Authors**
**Deepak Dagur -** CNR-IOM - Istituto Officina dei Materiali, S.S. 14 km 163.5, I-34149, Trieste, Italy and Department of Physics, University of Trieste, Trieste, Via Alfonso Valerio 2, 34127, Italy

**Alice Margherita Finardi -** Department of Physics, University of Milan, Milan, Via Festa del Perdono 7, 20122, Italy

**Vincent Polewczyk -** CNR-IOM - Istituto Officina dei Materiali, S.S. 14 km 163.5, I-34149, Trieste, Italy

**Aleksandr Yu. Petrov -** CNR-IOM - Istituto Officina dei Materiali, S.S. 14 km 163.5, I-34149, Trieste, Italy

**Simone Dolabella -** CNR-IOM - Istituto Officina dei Materiali, S.S. 14 km 163.5, I-34149, Trieste, Italy

**Federico Motti -** CNR-IOM - Istituto Officina dei Materiali, S.S. 14 km 163.5, I-34149, Trieste, Italy

**Hemanita Sharma** - CNR-IOM - Istituto Officina dei Materiali, S.S. 14 km 163.5, I-34149, Trieste, Italy and Department of Physics, University of Trieste, Trieste, Via Alfonso Valerio 2, 34127, Italy

**Edvard Dobovicnik -** Elettra - Sincrotrone Trieste S.C.p.A., Strada Statale 14 km 163.5, 34149 Trieste, Italy

**Andrea Giugni -** Department of Physics, University of Milan, Milan, Via Festa del Perdono 7, 20122, Italy

**Giorgio Rossi -** Department of Physics, University of Milan, Milan, Via Festa del Perdono 7, 20122, Italy

**Claudia Fasolato -** Istituto dei Sistemi Complessi (ISC)-CNR, Rome, Piazzale Aldo Moro 5, 00185, Italy

**Piero Torelli -** CNR-IOM - Istituto Officina dei Materiali, S.S. 14 km 163.5, I-34149, Trieste, Italy

**Author Contributions**

The manuscript was written through contributions of all authors. All authors have given approval to the final version of the manuscript.

**Notes**

The authors declare no competing financial interest.

ACKNOWLEDGMENT

This work was performed in the framework of the Nanoscience Foundry and Fine Analysis (NFFA-MUR Italy Progetti Internazionali) project ([www.trieste.NFFA.eu](www.trieste.NFFA.eu)).